\documentclass[pra,twocolumn,aps,showpacs]{revtex4-1}
\usepackage{amsmath,amssymb,graphicx}
\usepackage{natbib}
\usepackage{color}

\def\calh{{\cal H}}

\def\cali{{\cal J}}
\def\phi{\varphi}

\def\ff{\cal F}
\def\fp{\cal P}

\def\bc{\begin{center}}
\def\ec{\end{center}}

\begin{document}
\title{Fano resonances and band structure of two dimensional photonic structures }

\author{Peter Marko\v{s}}\email{Corresponding author: peter.markos@fmph.uniba.sk}
%\author[2,*]{Author Two}
%\author[1]{Author Three}

\affiliation{Department of Experimental Physics, Faculty of Mathematics, Physics and Informatics, Comenius University in Bratislava, 842 28 Slovakia
}

\begin{abstract}
We show that the frequency spectrum of two dimensional photonic crystals is strongly influenced by Fano resonances
which can be  excited already  in the linear array of dielectric cylinders. To support this claim, 
we calculate the transmission of electromagnetic wave through  linear  array of dielectric cylinders and show
that frequencies of observed Fano resonances coincides with position of narrow frequency bands found in the spectra of 
corresponding two-dimensional 
photonic crystals.  Split of frequency band or overlap of two bands,
observed in the band structure  of photonic structures are also associated with Fano resonances. 
\end{abstract}


\maketitle

\section{Introduction}

Frequency spectrum of photonic crystals consists of large number of continuous bands. 
This  is a direct consequence of spatial periodicity of the permittivity $\varepsilon$ which defines the structure.
In the limit of infinitesimally small variations of the permittivity,
 all  frequency bands  could be constructed from the dispersion relation of electromagnetic wave in homogeneous medium 
by reduction of the  momenta $\vec{q}$ to the first Brillouin zone \cite{sakoda,costas,iok}. With 
increasing permittivity contrast, frequency  gaps  open  at the edges of Brillouin zone \cite{joan-pc}.
We will call the resulting frequency bands periodic (${\fp}$) bands.

\begin{figure}[b!]
\bc
\includegraphics[width=0.64\linewidth]{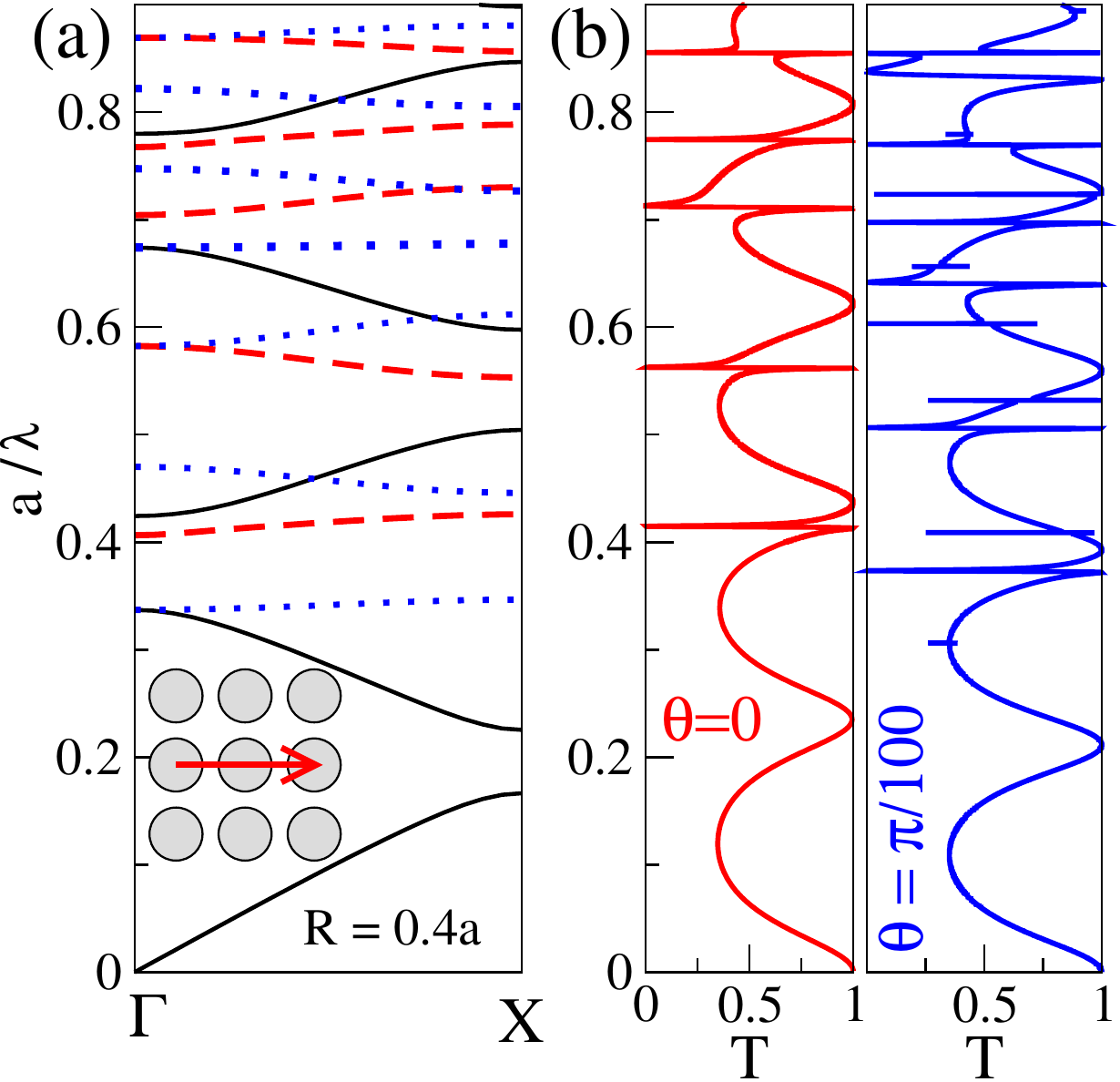}
~~
\includegraphics[width=0.3\linewidth]{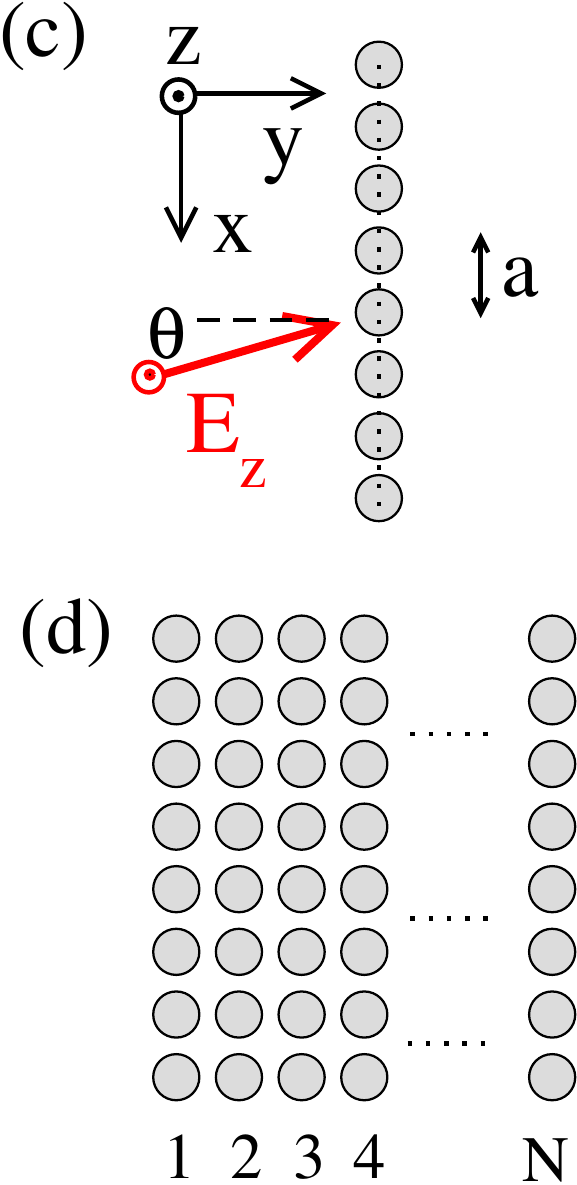}
\ec
\caption{(Color online) (a) Frequency spectrum of the square array of dielectric cylinders with radius $R=0.4a$
($a$ is the spatial periodicity in $x$ and $y$ directions)
and permittivity $\varepsilon=12$ calculated by the plane wave expansion method \cite{sakoda}. In the frequency region  $a/\lambda <0.9$ we found
 five  ${\cal P}$ bands (black full lines), five   even (dashed red lines) 
and seven odd   (dot  blue lines)  $\ff$ bands.
Inset shows schematically the structure in the $xz$ plane,  red arrow indicates the propagation direction of the electromagnetic wave.
(b) Transmission coefficient of the  plane wave incident on  the linear array  of cylinders 
(The geometry of the problem is shown in (c)).
The incident angle is $\theta=0$ (red line) and $\theta = \pi/100$ (blue line). 
Note the one-to-one correspondence between resonant frequencies of excited Fano resonances  
and  ${\cal F}$ bands observed in the frequency spectrum of the infinite 2D structure. 
(c) Transmission of incident electromagnetic wave through linear array of dielectric cylinders.
(d) Geometry of the photonic structure composed from $N$ rows of dielectric cylinders. ($N=24$ throughout this Paper.)
Both structures displayed in (c) and (d)  are  infinite in the $x$ direction.
}
\label{uvod}
\end{figure}

For higher permittivity contrast, the  frequency spectrum is more complicated.
It contains, besides the ${\fp}$ bands, also  other, usually very narrow (almost dispersionless) frequency bands. 
As an example, we show in Fig.  \ref{uvod}(a) the frequency spectrum of two dimensional 
square array of dielectric cylinders embedded in vacuum. Only five of 17 displayed frequency bands are $\fp$ bands.

The aim of this Paper  is to explain physical origin these additional frequency bands.  
We prove  that they originate from Fano resonances 
\cite{fano,miro}
which can be excited in  linear chain
of dielectric cylinders by  incident electromagnetic wave. 
We find numerically the spectrum
of these  resonances and demonstrate how each  resonance develops into narrow Fano ($\ff$) band
in corresponding two-dimensional photonic crystal.

The band structure shown in Fig. \ref{uvod}(a) was calculated for the square array of thick dielectric cylinders. For 
other structures, for instance photonic crystals composed from thinner cylinders, mutual coupling
of  $\ff$ and  original $\fp$ band occurs which results in irregularities of the band spectrum. Two of them, namely  
the split of the $\fp$ band, and an overlap of two bands, will be discussed later.

Fano resonances in photonic structures 
\cite{sfan} 
were studied  mostly  in process of interaction of 
two dimensional  photonic slabs with incident electromagnetic wave  
\cite{fan}
and were  used for experimental identification of spectra of leaky modes in photonic structures
\cite{astr}.
For the case of individual dielectric cylinder, 
Fano resonances were observed as a result of  the coupling of Mie resonant states \cite{mie,hulst}
with an  incident electromagnetic wave 
in spherical \cite{trib} and cylindrical dielectric object
\cite{rybin}.
Fano resonances play an important role in design of metamaterials 
\cite{rr,luk}
and influence considerably the  transport properties of  disordered systems
\cite{fano-nature}.
Here, we are interested  in Fano resonances which could be excited 
by plane electromagnetic wave incident to the linear periodic  array of dielectric cylinders (Fig. \ref{uvod}(c)) as a result of interference of leaky guided modes of periodic array of cylinders \cite{joan-pc} with incident electromagnetic wave. 
These resonances can be characterized by the resonant frequency and lifetime which is inversely proportional to  the width of the resonant peak. Surprisingly, they are narrower than resonances excited in individual dielectric cylinders.  If the electromagnetic wave  propagates across  $N$ chains of cylinders (Fig. \ref{uvod}(d)) then 
resonances excited in individual chains couple together and create narrow transmission band obtained in the frequency spectra of photonic crystals.

\section{Structure and method}

We concentrate on periodic  photonic structures composed from dielectric cylinders 
parallel to the $z$ axis. Cylinders  
possess a  frequency independent permittivity $\varepsilon=12$ and permeability $\mu = 1$. 
The embedding medium is vacuum. Radius of cylinders is $R$ and the distance between two nearest-neighbor cylinders
is $a$.

The first structure is the linear chain of dielectric cylinders
lying in the $y=0$ axis (Fig. \ref{uvod}(c)).  
Incident electromagnetic wave of wavelength $\lambda$  excites in this structure Fano resonances 
which manifests themselves as sharp irregularities in the frequency dependence of the transmission coefficient visible in
Fig. \ref{uvod}(b).
Transmission coefficient is calculated also for finite photonic slab constructed from $N=24$ rows of the same cylinders 
located in planes  $y = (n-1)a$ ($n=1,2,\dots,24$, see  Fig. \ref{uvod}(d)). Observed transmission spectra will be compared with 
the band structure of  an infinite square array of cylinders calculated by plane wave expansion method \cite{sakoda}.

Transmission coefficient  is calculated  by the  transfer matrix method \cite{pendry,pre}. 
For more detailed analysis of the spatial distribution of electric field, we use another algorithm based on the  expansion of 
electromagnetic field  into cylinder  functions \cite{hulst,stratton,on,oua}.
 If an incident electromagnetic wave is polarized  with electric field $E_z$ parallel to cylinders 
(this polarization is considered throughout this Paper) 
 then the  intensity of electric field 
scattered at a cylinder centered in $(xy)=(0,0)$ can be expressed as
\begin{equation}
\label{eq:inc}
\begin{array}{rclcl}
E_z^{\rm in}(r,\phi) &\!\!\!=\!\!\!& {\cali}_0(r)\alpha_{0}^+ 
&\!\!\!\!+\!\!\!& 2\displaystyle{\sum_{k>0} }\alpha_{k}^+{\cali}_k(r) \cos(k\phi)\\
&&& + &
 2i\displaystyle{\sum_{k>0}} \alpha_{k}^-{\cali}_k(r) \sin(k\phi)\\
E_z^{\rm out}(r,\phi) &\!\!\!=\!\!\!& {\calh}_0(r)\beta_{0}^+ 
&\!\!\!\!+\!\!\!& 2\displaystyle{\sum_{k>0} }\beta_{k}^+{\calh}_k(r) \cos(k\phi)\\
&&& + &
 2i\displaystyle{\sum_{k>0}} \beta_{k}^-{\calh}_k(r) \sin(k\phi)
\end{array}
\end{equation}
for $r<R$ and $r>R$, respectively. Similar expressions for radial and tangential components of magentic field could be derived from Maxwell equations \cite{stratton}. 
In Eq. (\ref{eq:inc}),
${\cali}_k(r) = J_k(2\pi n r/\lambda)$,
${\calh}_k(r) = H_k(2\pi r/\lambda)/H'_k(2\pi R/\lambda)$, \cite{pozn}
 $J_k$, $H_k$  are Bessel and Hankel function, 
$\lambda=2\pi c/\omega$ determines the wavelength of electromagnetic field in vacuum and 
$n=\sqrt{\varepsilon\mu}$ is the index of refraction.

For another cylinder, centered  at $x=n_xa$ and $y=n_ya$ 
the field $E_z$ is again expressed by Eq. \ref{eq:inc} but 
with new set of coefficients $\alpha(n_x,n_y)$,  $\beta(n_x,n_y)$   and  
cylindrical  coordinates  $r$ and $\phi$  associated with the center of the cylinder.

In numerical simulations, we calculate coefficients  $\alpha$ and $\beta$ from the requirement of continuity of tangential components  of the intensity of electric and magnetic field at the boundary of cylinders. 
Note that spatial periodicity of the structure along the $x$ direction 
considerably reduces the number of unknown coefficients since 
coefficients $\alpha(n_x,n_y)$ and  $\beta(n_x,n_y)$    fulfill the Bloch theorem
\begin{equation}
\label{eq:bloch}
\alpha(n_x,n_y) = \alpha(0,n_y)e^{iqan_x},~~~~ 
\beta(n_x,n_y) = \beta(0,n_y)e^{iqan_x}
\end{equation}
where $q$ is the  transverse component of the wave vector. This enables us to reduce the number of unknown coefficients
$\beta^\pm_k(n_y)$
to $N\times (2N_B+1)$, where $N_B$ is the highest order of Bessel function used in numerical calculations ($N_B=12$ in most cases).
%\footnote{The file on the outer boundary of cylinder is a superposition of waves scattered on all other cylinders.}
To find the transmission coefficient $T$,  Poynting vector is calculated on the opposite side of the structure. Details of the method are given elsewhere \cite{pm-15}. 

\begin{figure}[t!]
\bc
\includegraphics[width=0.9\linewidth]{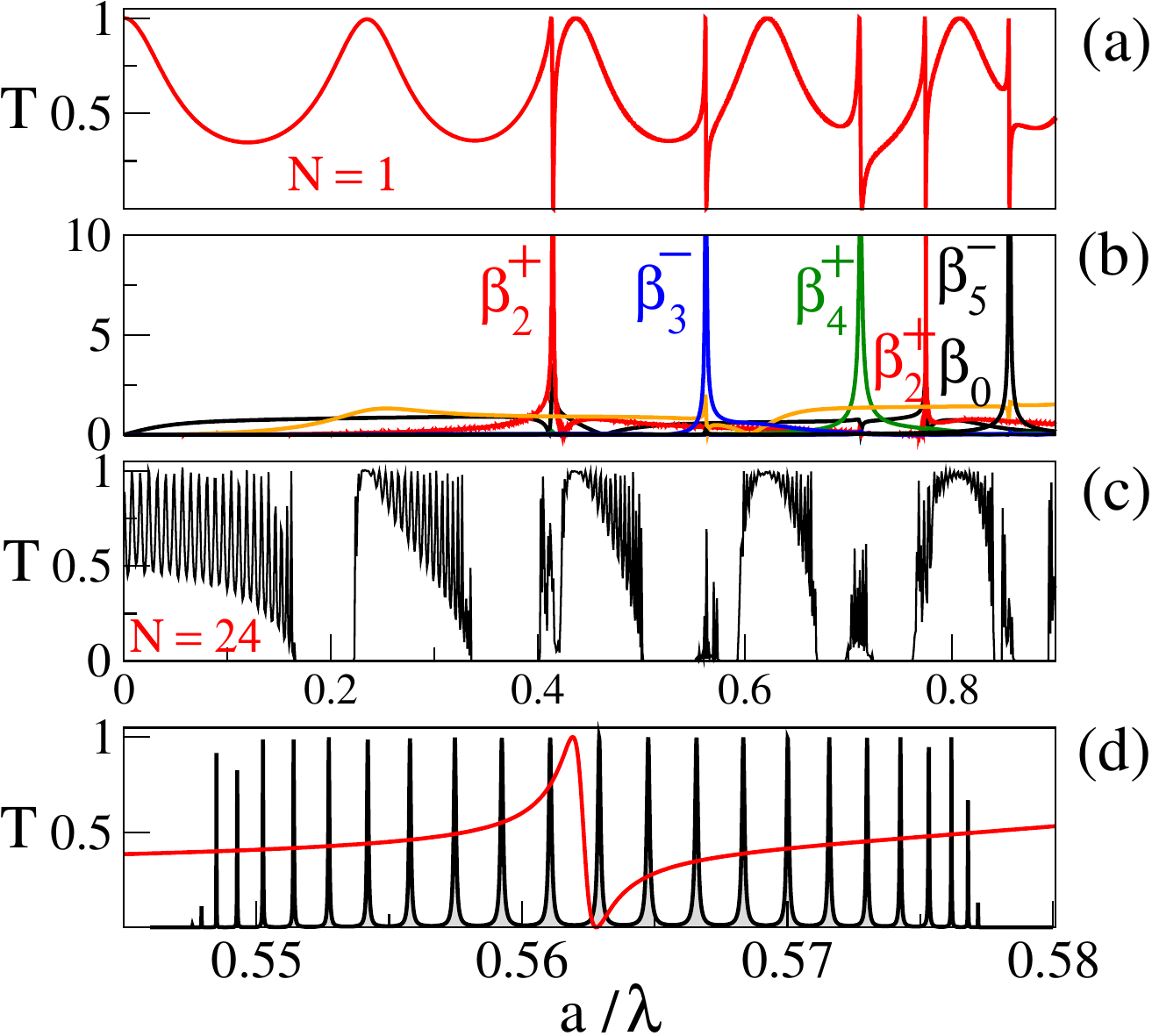}
\ec
\caption{(Color online) (a)  The transmission coefficient  of the linear chain of cylinder with radius $R=0.4a$
shown in Fig. \ref{uvod}(c).
Only even resonances are excited since  the electromagnetic wave propagates perpendicularly  
to the cylinder chain \cite{pc-asym,sak}.
(b) The frequency dependence of parameters $\beta_k^\pm$.
%The  $k^+$ ($k^-$)  resonance corresponds to the $k$th cosine (sine) mode in Eq. \ref{eq:inc}.  
(c) Transmission coefficient for $N=24$ rows of cylinders. Fano resonances observed in linear chain develop 
themselves to narrow frequency bands.
(d) Detail of the the frequency dependence of the transmission coefficient shown in (a) and (c) 
in the vicinity of the  Fano resonance of coefficient $\beta_3^-$  ($a/\lambda= 0.562$).
}
\label{r40}
\end{figure}

\begin{figure}[t]
\begin{tabular}{lccccc}
${\fp}$ bands: &1st & 2nd & 3rd & 4th & 5th \\
~ & ~
\includegraphics[width=0.11\linewidth]{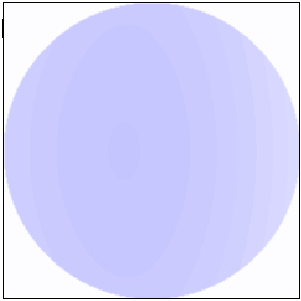}
~& ~
\includegraphics[width=0.11\linewidth]{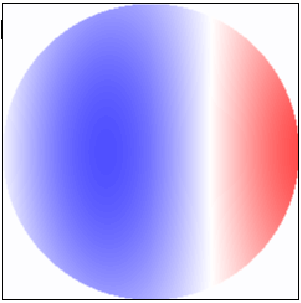}
~& ~
\includegraphics[width=0.11\linewidth]{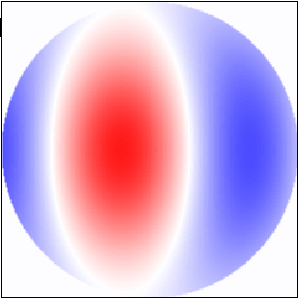}
~& ~
\includegraphics[width=0.11\linewidth]{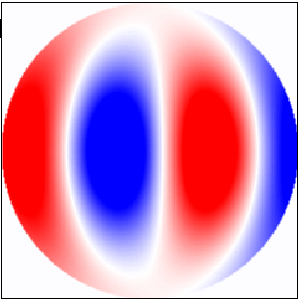}
~& ~
\includegraphics[width=0.11\linewidth]{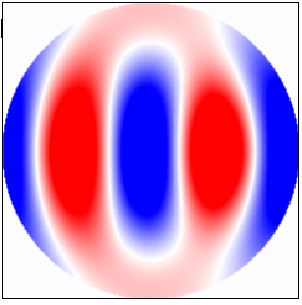}
~ \\
${\ff}$ bands: & $2^+$ & $3^-$ & $4^+$ & $0,2^+$ & $5^-$\\
~&~
\includegraphics[width=0.11\linewidth]{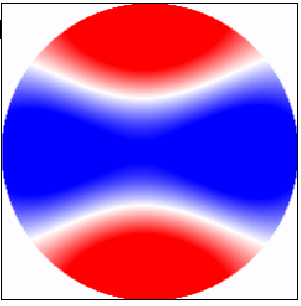}
~&~
\includegraphics[width=0.11\linewidth]{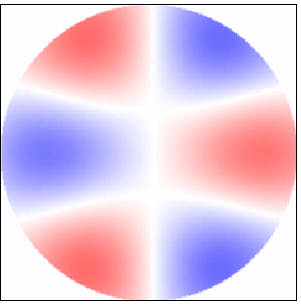}
~&~
\includegraphics[width=0.11\linewidth]{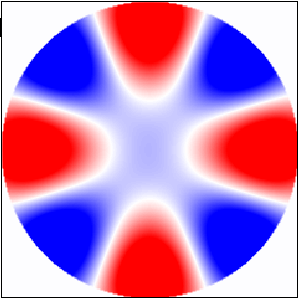}
~&~
\includegraphics[width=0.11\linewidth]{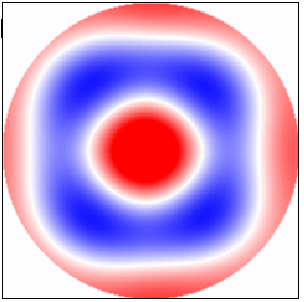}
~&~
\includegraphics[width=0.11\linewidth]{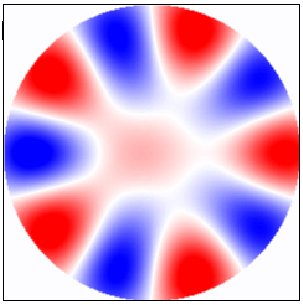}
~
\end{tabular}
\caption{(Color online) Comparison of the  field distribution $E_z$ in dielectric cylinder for
frequencies in the center of  the  five lowest $\fp$ bands (top)
and five lowest ${\cal F}$ bands (bottom). Electromagnetic wave propagates through photonic slab of the thickness of 24 rows of cylinders.
For each frequency, only field inside the $10$th cylinder along the $y$ direction is shown.
}
\label{r40-fano}
\end{figure}

%%%%%%%%%%%%%%%%%%%%%%%%%%%%%%%%%%%%%%%%%%%%%%%

\section{Results}

\begin{figure}[b!]
\bc
\includegraphics[width=0.232\linewidth]{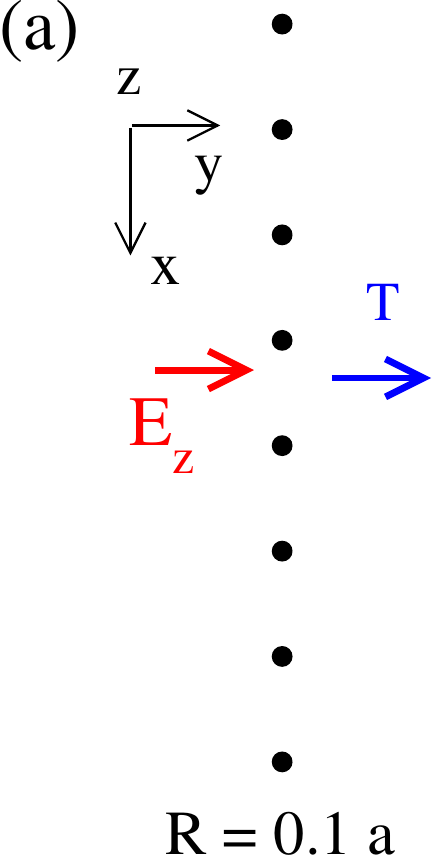}
~~
\includegraphics[width=0.72\linewidth]{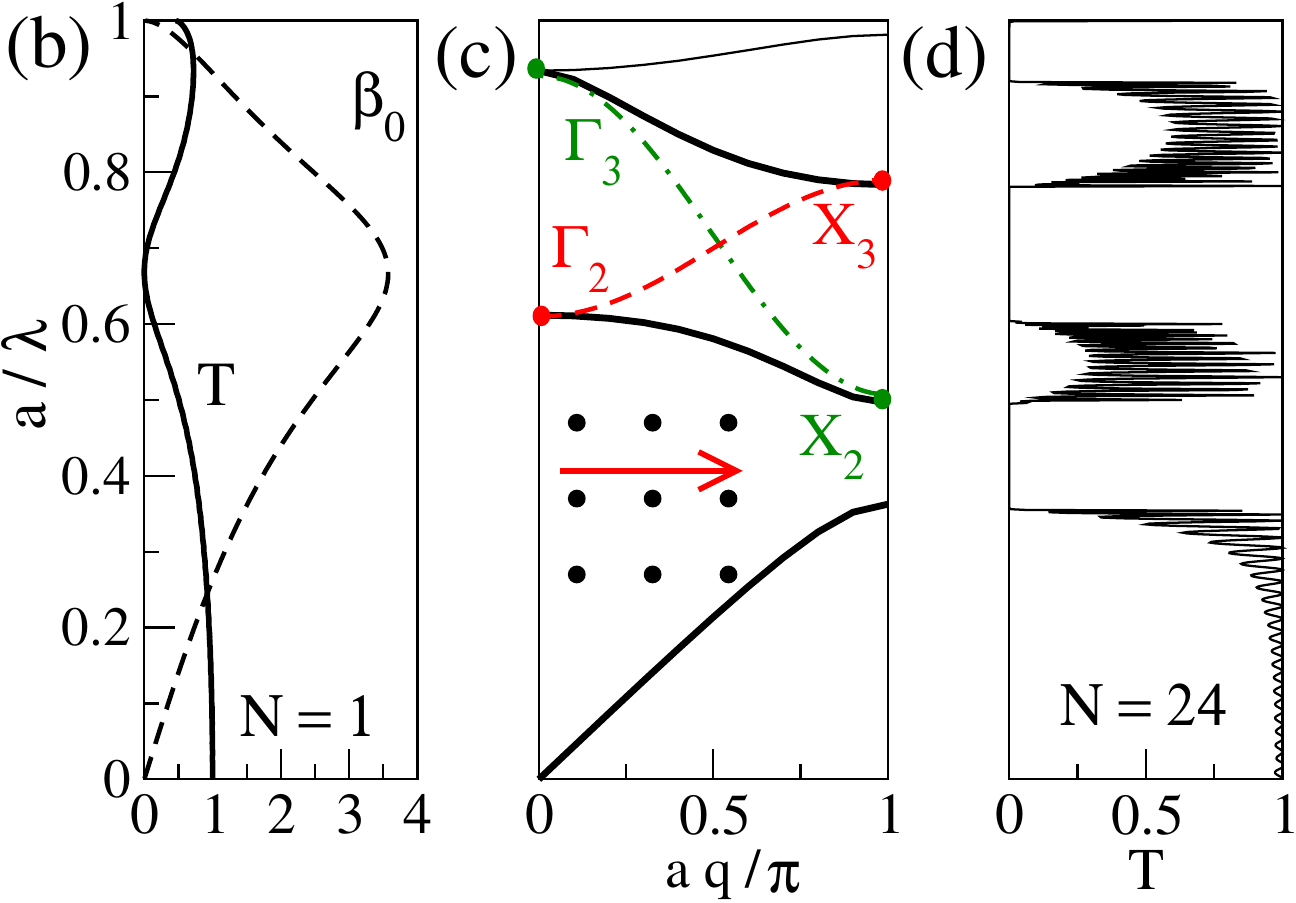}
\ec

\noindent{(e)}

\bc
\textcolor{green}{$\Gamma_3$}~~~\includegraphics[width=0.9\linewidth,height=0.04\linewidth]{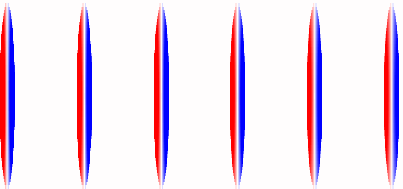}
\textcolor{red}{X$_3$}~~~\includegraphics[width=0.9\linewidth,height=0.04\linewidth]{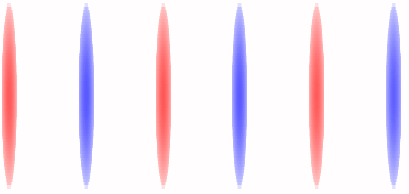}
\textcolor{red}{$\Gamma_2$}~~~\includegraphics[width=0.9\linewidth,height=0.04\linewidth]{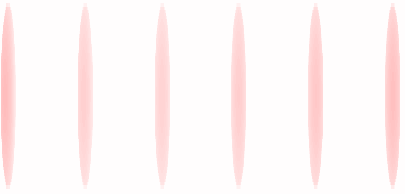}
\textcolor{green}{X$_{2}$}~~~\includegraphics[width=0.9\linewidth,height=0.04\linewidth]{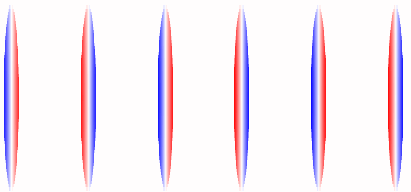}
$\longrightarrow~~y$
\ec
\caption{(Color online) (a) Linear chain of thin dielectric cylinders ($R=0.1a$) and 
(b)  Transmission coefficient $T$  and the coefficient $\beta_0$ as a function of the frequency $a/\lambda$.
Owing to  Fano resonance the transmission decreases to zero at $a/\lambda\approx 0.67$.
(c)  The band structure of two-dimensional square array. Fano resonance causes the broad gap between the 
second and the third bands. 
(d) The transmission coefficient of the electromagnetic wave propagating through an array of 24 rows of
cylinders. (e) The intensity of electric field  $E_z$ inside six cylinders along the $y$ direction at band edges 
 X$_{2,3}$ and $\Gamma_{2,3}$. 
}
\label{r10}
\end{figure}

\begin{figure}[t]
\bc
\includegraphics[width=0.4\linewidth]{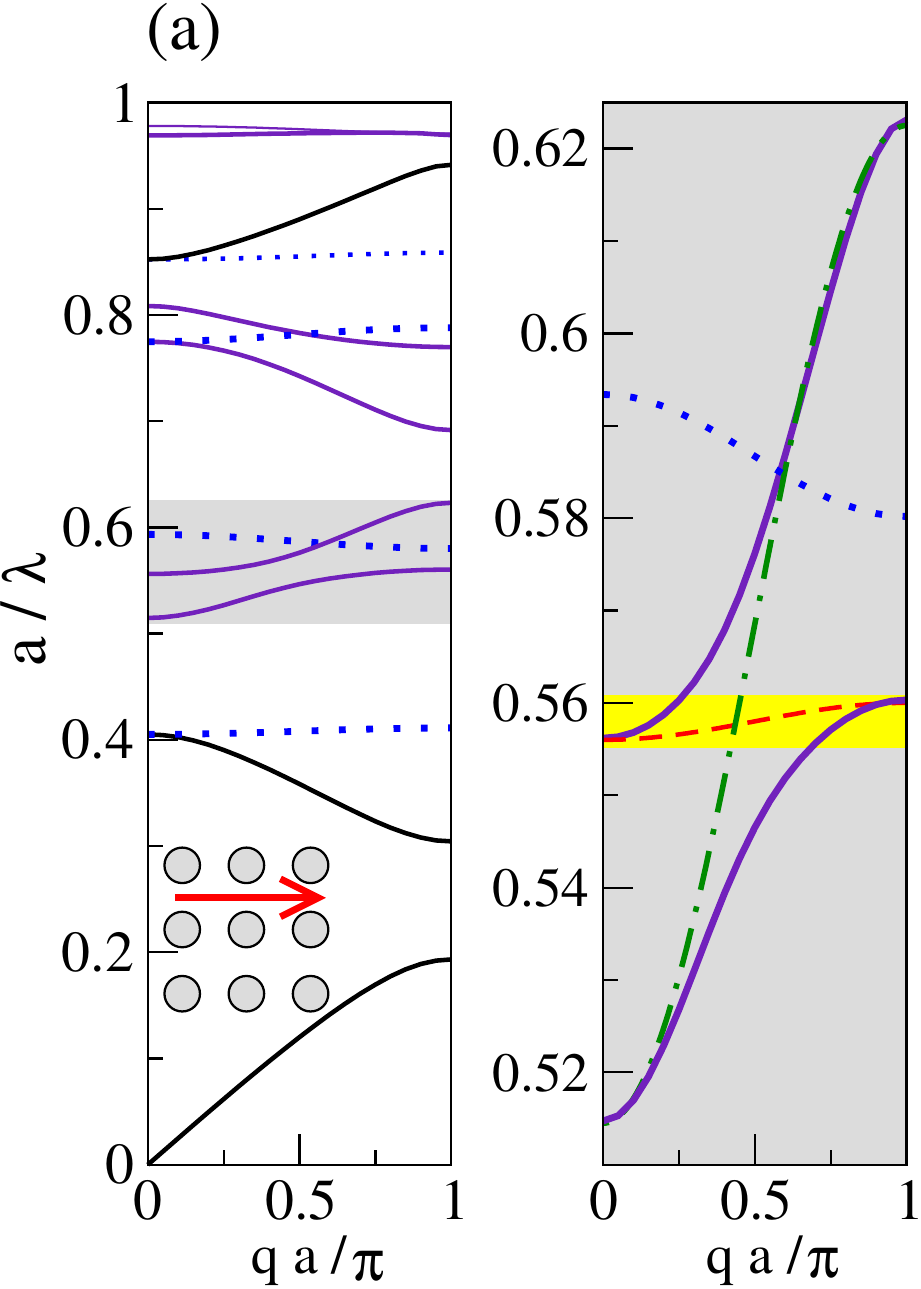}
~~~~
\includegraphics[width=0.50\linewidth]{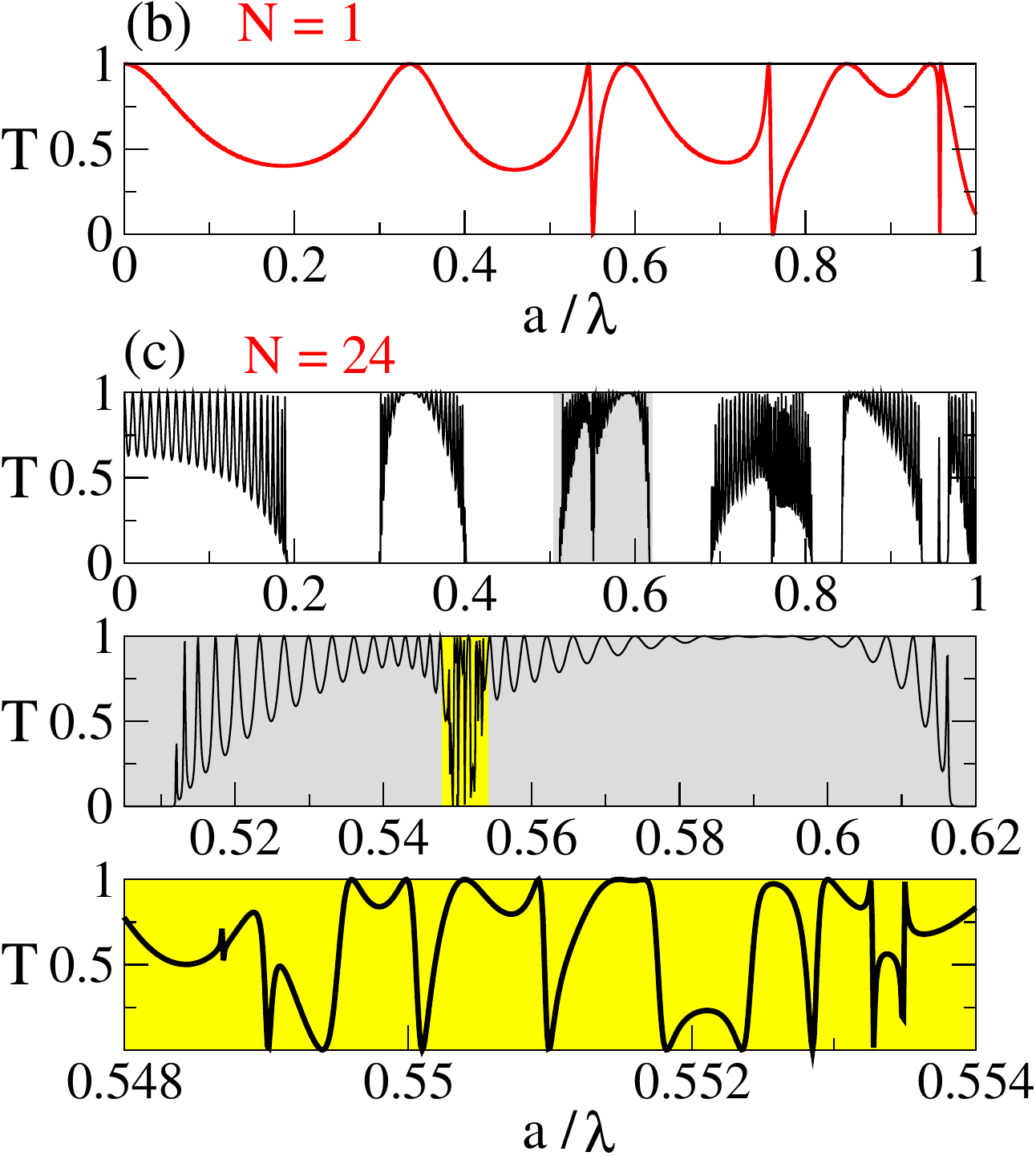}
\ec
\caption{(Color online) 
(a) Frequency spectrum of a square array of dielectric cylinders with radius $R=0.3a$. 
Solid lines represent bands with even spatial symmetry, 
dot lines correspond, similarly to Fig. \ref{uvod}(a), to odd Fano bands.
Note an overlap of the 3rd and 4th bands, shown in detail in the
right panel. This overlap  results from the  coupling of 
two bands displayed by green dot-dashed and red dashed line. 
(b) Frequency dependence of the transmission coefficient for the linear chain of cylinders. 
Three Fano resonances could be easily identified.
(c) Top panel shows the transmission coefficient for 24 rows of cylinders  as a function of
frequency $a/\lambda$. Two lower panels present detail of the  frequency dependence of the transmission coefficient in the vicinity of the Fano resonance. \cite{xx}
}
\label{eps12-T}
\end{figure}

\subsection{Isolated Fano bands}

Consider first  an array of  thick dielectric cylinders with radius $R=0.4a$.  
Figures \ref{uvod}(b) and \ref{r40}(a)  shows the transmission coefficient of plane electromagnetic wave propagating 
through linear chain of cylinders. 
A series of very narrow resonances could be identified. Similar maxima and minims has been found numerically in the reflection coefficient  \cite{on}.
Following \cite{fan} we interpret these resonances as Fano resonances which results from the interference of incident plane wave with 
leaky guided modes excited in the periodic cylinder row \cite{joan-pc}. Indeed, 
these Fano resonances are accompanied by sharp maxima of coefficients $\beta$ defined in Eq. \ref{eq:inc}  (Figure \ref{r40}(b)). 
More detailed analysis of resonances is given in Sect. \ref{sect:fano}.

Comparison of  Figures \ref{uvod}(a) and (b)  confirms that Fano resonances
develop to narrow Fano bands in the spectra of 2D structures.
This is also shown  in Figs. \ref{r40}(c,d) 
which present the frequency dependence of the transmission coefficient of plane wave propagating through the  slab composed  from $N=24$ rows of cylinders.

Different character  of $\fp$ and $\ff$  bands  is clearly visible from the spatial symmetry of the 
electric field $E_z$ shown  in 
Fig. \ref{r40-fano}.
The top panel displays  the field $E_z$ for frequencies chosen in the center of 
five lowest ${\fp}$ band shown in Fig. \ref{r40}(c). As  expected, 
the field symmetry changes when the frequency increases from one frequency band to the next one \cite{sakoda,joan-pc}. 
The bottom panel shows the field $E_z$ for the five lowest 
resonant frequencies identified in Fig. \ref{r40}(b). As shown in Fig. \ref{r40}(d), these frequencies correspond to the center
of $\ff$ bands. The symmetry of the field is unambiguously determined  by the order of excited  Fano resonance.

\subsection{Overlap of two bands}

For thinner cylinders, the   $\fp$ band and the $\ff$ band can overlap. Then, the  resulting frequency spectrum depends on the $q$-dependence  of the frequency in two bands  
and on the strength of their mutual coupling. 
Consider a simple model of two bands $2V_1\cos q$ and $2V_2\cos q$, (centered at the same frequency for simplicity)  coupled together with coupling constant $2A$.
Resulting spectrum has a form
\begin{equation}
\omega_{1,2}(q) = (V_1+V_2)\cos q \pm \sqrt{|A|^2+(V_1-V_2)^2\cos^2q}
\end{equation}
 Two bands $\omega_1(q)$ and $\omega_2(q)$ are separated by gap if $4V_1V_2<|A|^2$.
 This happens either when $|A|$ is large or when one of two bands is narrow. 

\bigskip

%\subsubsection{Band splitting}
\noindent\textsl{Band splitting}. 
We found the above mentioned band  splitting in the frequency spectrum of the 
square array of thin   ($R=0.1a$) dielectric cylinders
(Fig. \ref{r10}).
The transmission coefficient through a linear chain of cylinders
(Fig. \ref{r10}(b))
decreases to zero for the frequency $a/\lambda\approx 0.67$. This decrease is 
accompanied by an increase of coefficient $\beta_0$ (Eq. \ref{eq:inc}).
We interpret this decrease of the transmission as a result of excitation of  broad Fano resonance.

Figure \ref{r10}(c)  shows the band structure of the square array of thin cylinders. 
Note that both the second and the third bands have a minimum  at the X point. 
These two bands result  from the coupling of the  ${\cal P}$  band  with broad  ${\cal F}$ band
displayed by dot dashed (X$_2\Gamma_3$) and dashed  ($\Gamma_2$X$_3$) lines, respectively. 

This statement is supported also by the analysis  of the spatial symmetry of electric field  within two bands
shown in Fig. \ref{r10}(e). Note that the symmetry of electric field changes  along the line X$_2\Gamma_2$.
Also, the field in points X$_2$ and  $\Gamma_3$ have the same symmetry; of course,  the same holds for pair X$_3$ and $\Gamma_2$.
While the field at X$_2$ and $\Gamma_3$ possesses 
the symmetry of the ${\cal P}$ band, field close to inner band edges X$_3$ and $\Gamma_2$ has a symmetry of excited Fano resonance.

\bigskip

%\subsubsection{Band overlap}
\noindent\textsl{Band overlap}.
The  overlap of $\fp$ and $\ff$  bands is observed in 
 the band structure of the infinite square array of  cylinders with  radius  $R=0.3a$
displayed in Fig. \ref{eps12-T}(a) 
As shown in the right panel, the overlap of the 3rd and 4th bands  can be interpreted 
 as a result of coupling of two bands shown by dot dashed and dashed lines, respectively.  
This  overlap can be identified also from the 
complicated frequency dependence of the transmission coefficient \cite{sakoda-1997} shown in Fig. \ref{eps12-T}(b)
for the slab of $N=24$ rows of cylinders.  
Similar analysis could be done for the overlap of the 5th and 6th bands in Fig. \ref{eps12-T}(a).\cite{xx}

\section{Fano  resonance}\label{sect:fano}

In previous Section, we have shown that  Fano resonances excited in linear array of dielectric cylinders create the $\ff$ bands 
in spectra of photonic crystals. Now we will discuss physical origin of Fano resonances.

Fano resonances have been observed recently in the most  simple dielectric structure 
-- the single dielectric cylinder \cite{rybin}.
If the frequency of incident electromagnetic wave coincides with  the 
eigenfrequency of any cylinder leaky eigenmode,   the last can be  excited.
Then,  the electromagnetic field in the neighbor of the cylinder
is given by a  superposition of two fields  with the same frequency: the incident plane wave and 
field radiated by excited resonance \cite{rybin}. The excitation of resonance manifests itself as a maximum of coefficient $\beta$ shown in Fig. 6(a). The width of the resonance is proportional to inverse of its lifetime. 
Similarly, excitation of resonant  guided mode in 
linear array of cylinders can be identified from   sharp maxima in frequency dependence of corresponding coefficient $\beta$ (Figs. 2(b), 6(b)).
The interference of two modes is the responsible for  narrow maxima and minima in the transmission coefficient  of incident electromagnetic wave 
displayed for instance in  Figs. 1(b) and  2(a).

\begin{figure}[t]
\bc
\begin{minipage}[b]{0.26\textwidth}
\includegraphics[width=0.99\linewidth]{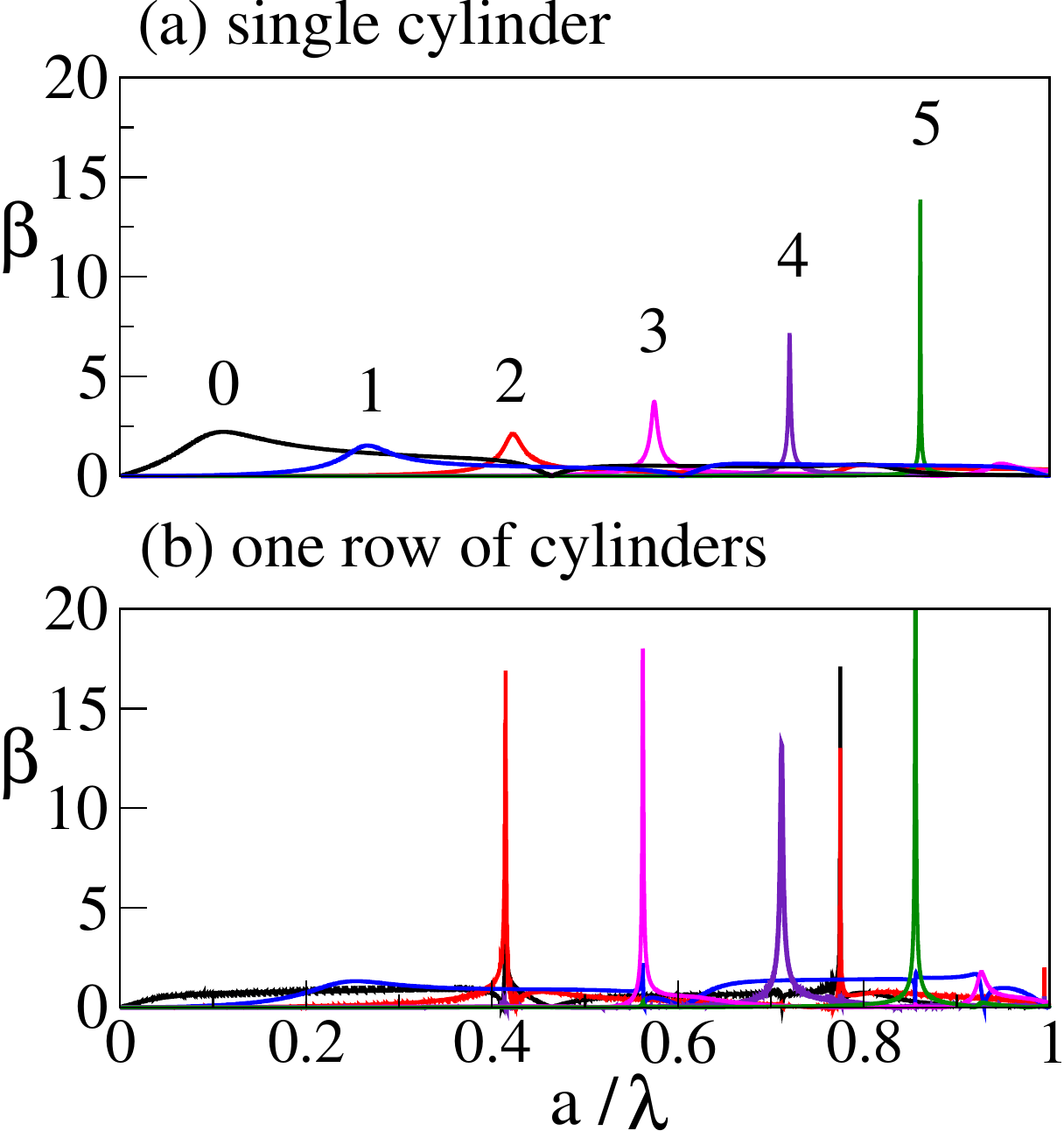}\\[0.5cm]
\includegraphics[width=0.89\linewidth]{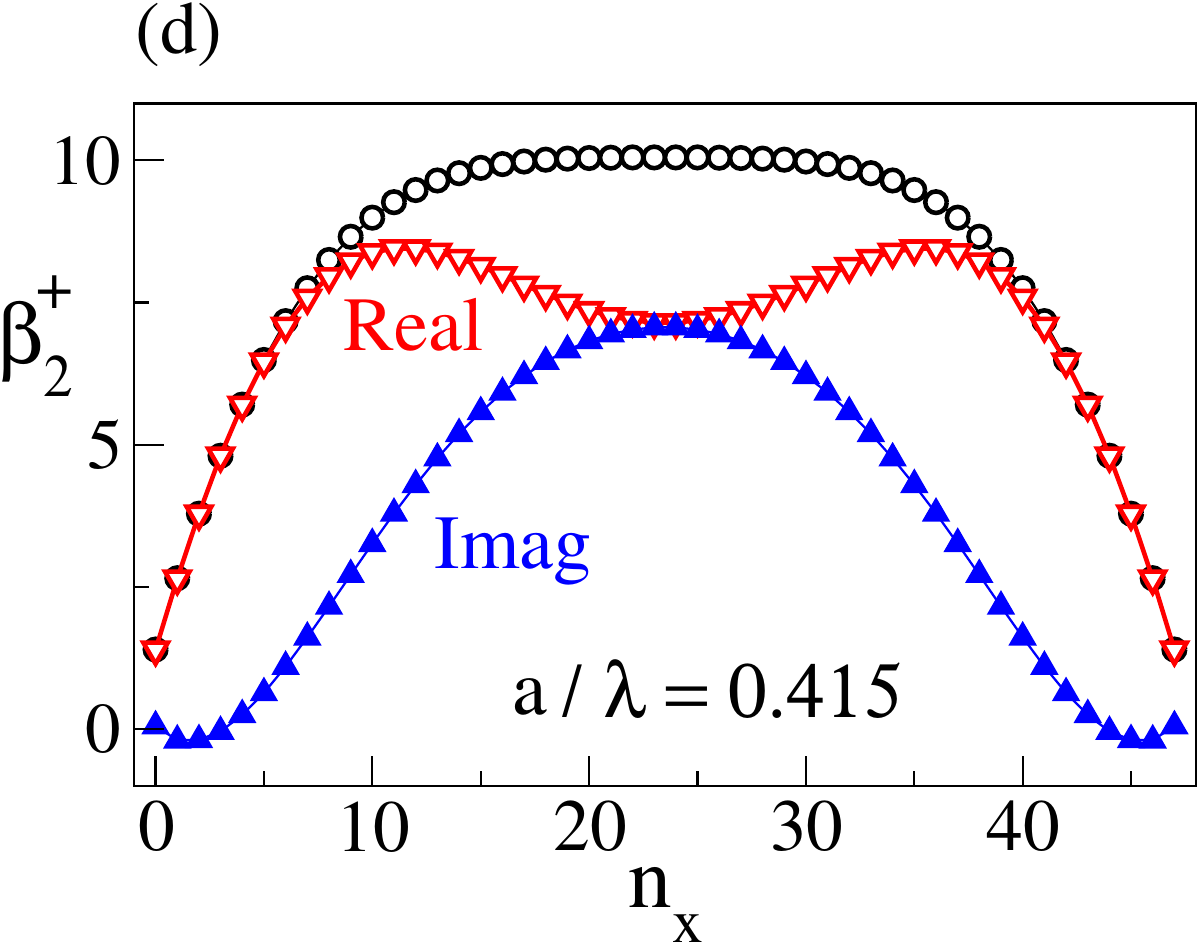}
\end{minipage}~~\begin{minipage}[b]{0.22\textwidth}
\includegraphics[width=0.99\linewidth]{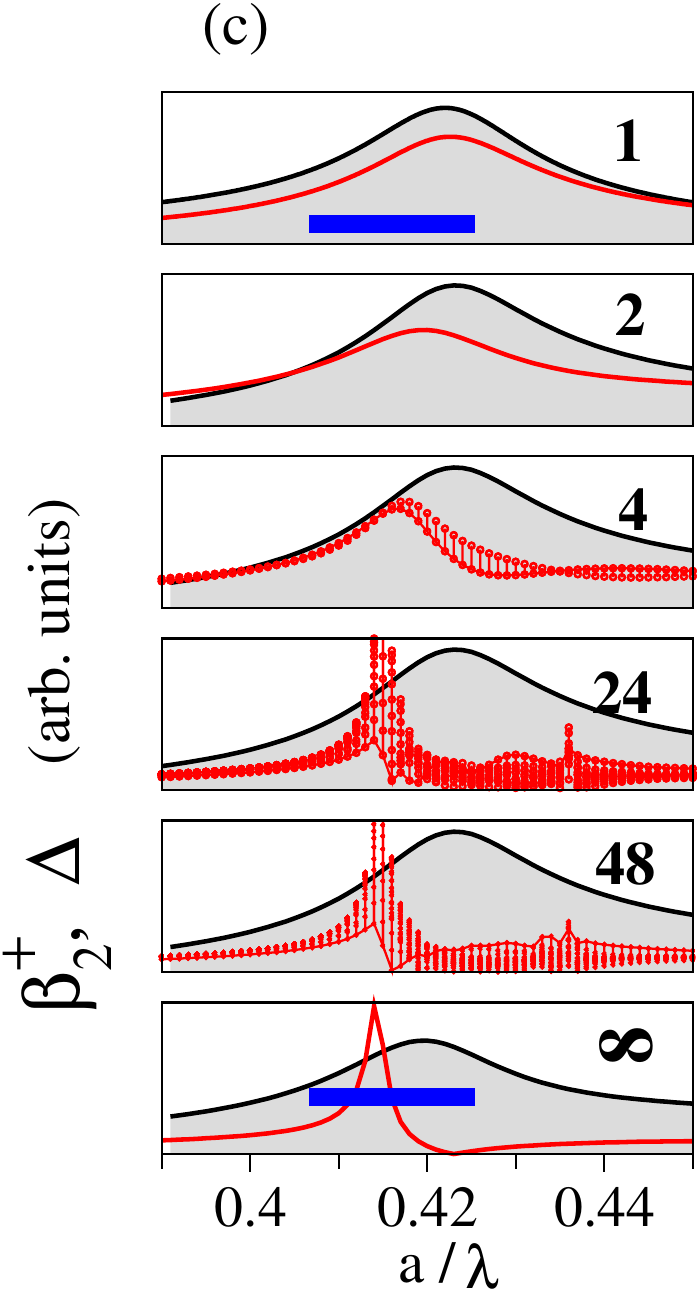}
\end{minipage}
\ec
\caption{(Color online) (a) Coefficients $\beta$ for  (a) single dielectric cylinder  with radius $R=0.4a$.
(b)   linear row of cylinders.   (c)  Coefficients  $\beta_2^+$ (red) and an inverse determinant $\Delta^{-1}$ (black) for the  clusters of $M$ dielectric cylinders ($M=1,2,4,24, 48,\infty$) along the $x$ axis. Thick blue line 
shows the position of the $\ff$ band $0.4067 < a/\lambda < 0.4254$ displayed in Fig. 1(a) by red dashed line. 
Multiple points for a given frequency corresponds with   different intensity of excited resonance
 at individual cylinders  as shown in (d) for the cluster with  $M=48$ cylinders.
}
\label{fig:mie}
\end{figure}

We start with the comparison  of the frequency dependence of 
 coefficients $\beta$ for isolated dielectric cylinder 
 (Fig. \ref{fig:mie}(a) ) 
and for an infinite periodic array of cylinders 
(Figs. \ref{r40}(b)  and \ref{fig:mie}(b)).  
For single cylinder and $E_z$ polarization, $\beta_k$ is a solution of system of linear equations \cite{stratton}
\begin{equation}
\left(
\begin{array}{cr}
{\cal J}_k & - {\cal H}_k(R) \\
{\cal J'}_k & - \zeta  
\end{array}
\right)
\left(
\begin{array}{c}
\alpha_k\\
\beta_k
\end{array}
\right)
=
\left(
\begin{array}{r}
J_k \\
\zeta J'_k
\end{array}
\right)
\end{equation}
where $J_k=J_k(2\pi R/\lambda)$, 
$\zeta=\sqrt{\mu/\varepsilon}$ 
is an impedance,
and the r.h.s is given by the expansion of
incident plane wave into Bessel functions \cite{as}
\begin{equation}
e^{i 2\pi R/\lambda \sin\theta} = \sum_k J_k(2\pi  R/\lambda)e^{ik\theta}
\end{equation}
As shown in Fig. \ref{fig:mie}(a), Fano  resonances of cylinder lye very close to those of  an array of cylinders  (Figs. 2(b) 
and Fig. \ref{fig:mie}(b)). The first two resonances ($k=0$ and $k=1$ are relatively broad, especially 
for an infinite number of cylinders.

On the other hand,  Fano resonances with $k\ge 2$ 
are significantly  narrower
when excited in an  infinite linear chain of cylinders 
than   in  individual cylinder. To explore how the shape of the resonance depends on the number of cylinders, we 
analyze the scattering of incident electromagnetic wave on finite cluster consisting from $M$ 
cylinders along the $x$ direction (Fig. 1(c)).
Coefficients $\beta_k^\pm(n_x)$  were calculated as a solution of system of linear equations
\begin{equation}
\label{eq:lin}
\textbf{A} \vec{\beta} = \vec{a}
\end{equation}
where $\textbf{A}$ is a matrix of the size $M\times (2N_B+1)$  and vector $\vec{a}$ represents incident electromagnetic wave.
Figure \ref{fig:mie}(c)   presents coefficients $\beta_2^+$ for finite cluster of $M$ cylinders excited  by perpendicularly incident 
plane wave. The resonance indeed becomes narrower when number of cylinders increases. Since the spatial distribution of electromagnetic field along the finite cluster is not homogeneous, $\beta_2^+$ acquires various values for individual cylinders inside the cluster. As an example, we plot in Fig. \ref{fig:mie}(d) spatial distribution of $|\beta_2^+|$ as well as its real and imaginary part calculated in cluster of 48 cylinders.

Figure \ref{fig:mie}(c) shows also the frequency dependence of an inverse of the determinant of the matrix $\textbf{A}$ (Eq. \ref{eq:lin})
which determines the eigenfrequencies  and lifetimes  of leaky guided modes \cite{economou} exited in the cluster.
Comparison of the frequency dependence of an inverse determinant and $\beta_2$ confirms   Fano character of observed excitation
and enables to estimate the sign of the Fano parameter $q<0$ \cite{miro,p2}.

\section{Absorption}

\begin{figure}[t]
\bc
\includegraphics[width=0.55\linewidth]{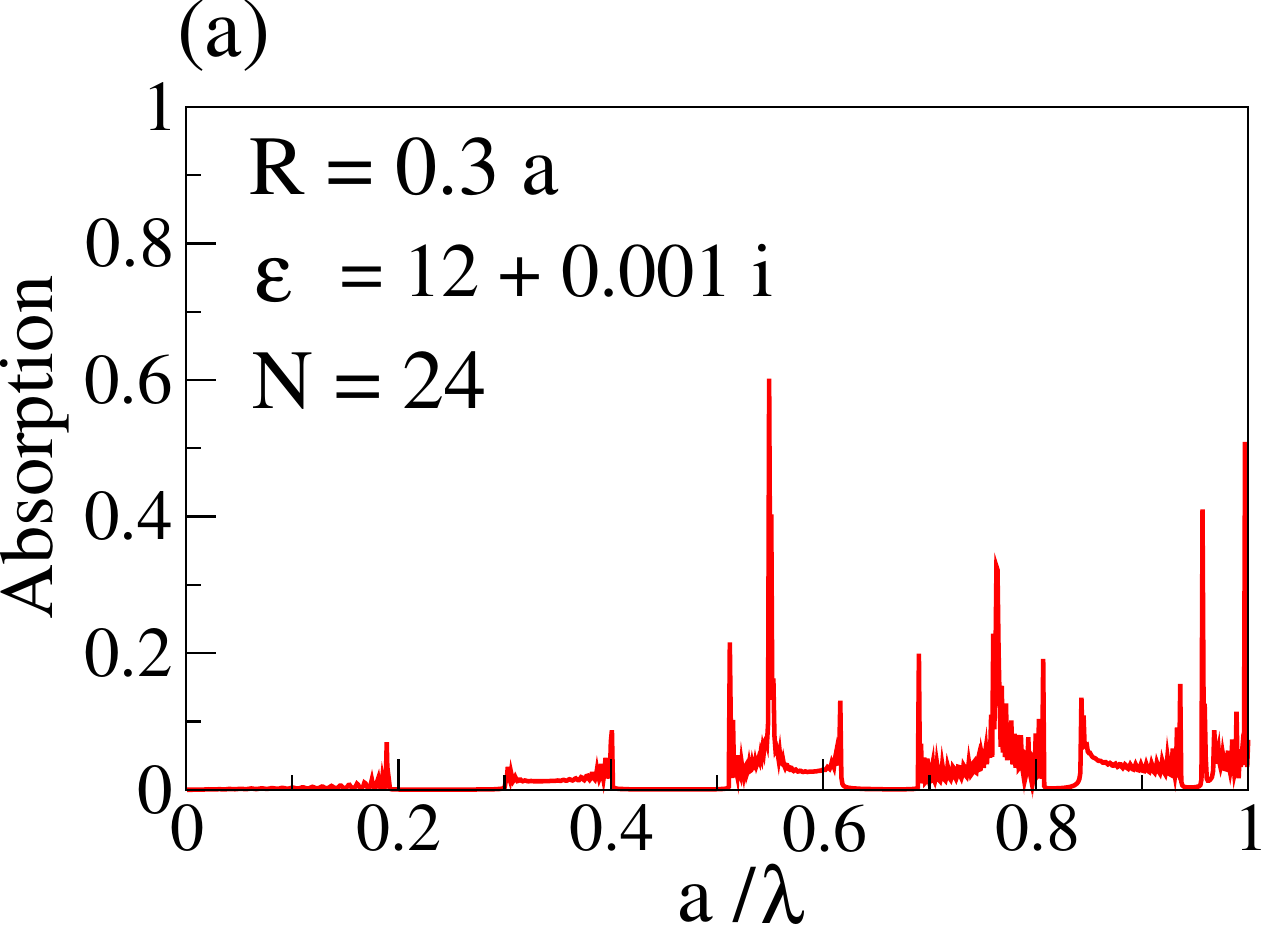}\\
~~~\\

\includegraphics[width=0.57\linewidth]{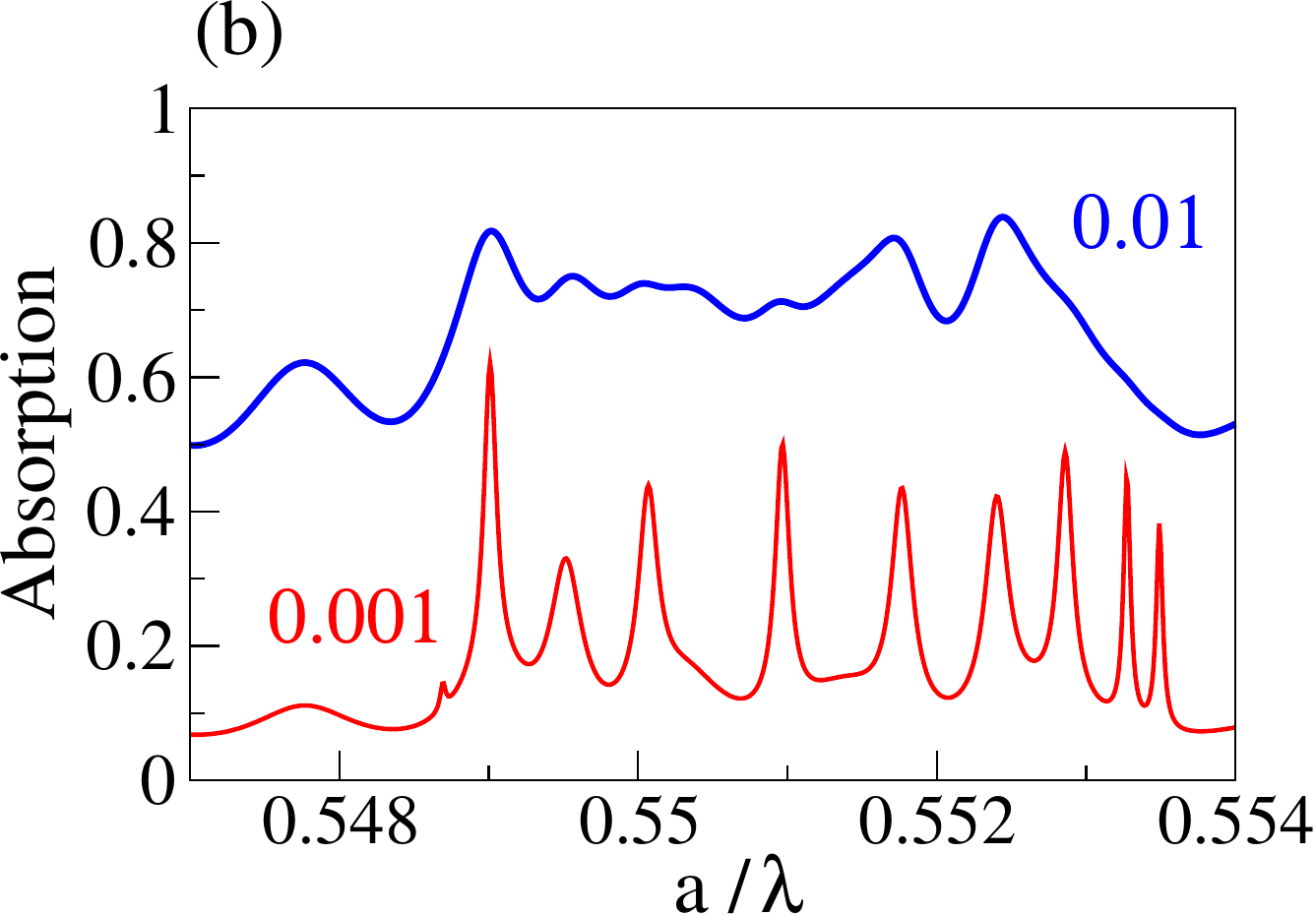}
\ec

\caption{(Color online) (a) Absorption of the electromagnetic wave in the photonic structure discussed in Fig. \ref{eps12-T} but  with  small imaginary part of the cylinder permittivity.
Absorption is large  at band edges, where the group velocity is small, and in the region of the Fano bands. 
(b) Detailed frequency dependence of the absorption in the resonant frequency region for
Imag $\varepsilon= 0.001$ and 0.01.
}
\label{abs}
\end{figure}

Finally, we  note  an another difference between $\fp$ and $\ff$ bands:  
we expect   that the  $\ff$  bands are much more sensitive to the absorption loses than the $\fp$ bands. 
One reason is  that  typical $\ff$  band is  narrow, therefore 
the group velocity of transmitted wave is small.  However, more important is that $\ff$ bands are associated with the resonance  which led to higher  intensity of propagating electric
field. Figure \ref{abs} presents the  absorption of electromagnetic field 
in the array of $R=0.3a$ cylinders with small imaginary part of the permittivity. 
Fano bands could be identified from the position of large maxima of the absorption. 
Note that very small imaginary part of the permittivity (Imag $\varepsilon$/Real $\varepsilon= 8.3\times 10^{-5}$)
causes the absorption of 20\% of energy when wave propagates through an array of 24 rows of cylinders.

\section{Conclusion}

In conclusion, we showed that the band structure of square arrays of cylinders can be completely described in terms of two kinds of frequency bands.
The $\fp$ bands originates from the reduction of the dispersion relation to the first reduced zone. 
The $\ff$ bands have origin in Fano resonances
observed  in the linear chain of cylinders. The field distribution in $\ff$ bands is determined by the symmetry of Fano resonance.
Two bands, $\fp$ and $\ff$ might overlap, which       complicates resulting band structure. 
Since resonance is typically accompanied by strong electric field, we expect that absorption is stronger in $\ff$ bands than in the $\fp$ bands.

\medskip

This work was supported by the Slovak Research and Development Agency under the contract No. APVV-0108-11
and by the Agency  VEGA under the contract No. 1/0372/13.


\begin{thebibliography}{99}

\bibitem{sakoda} K.~Sakoda,  \textsl{Optical Properties of Photonic Crystals}, Berlin, Heidelberg: Springer (2005).

\bibitem{costas} C.~M.~Soukoulis (Editor), \textsl{Photonic Crystals and Light Localization in the 21st Century}, NATO Sci. Ser. C.: Mathematical and Physical Sciences \textbf{563} Kluwer Acad.  Publ. (2001).

\bibitem{iok} K.~Inoue and K.~Ohtaka (Editors), \textsl{Photonic Crystals: Physics, Fabrication and Application}, Springer (2010).

\bibitem{joan-pc} J.~D.~Joannopoulos, S.~G.~Johnson,   J.~N.~Winnand  R.~G.~Meade, \textsl{Photonic Crystals: Molding the Flow of Light}  2nd edition. Princeton: Princeton University Press (2008).

\bibitem{fano} U.~Fano, Phys. Rev. \textbf{124}, 1866 (1961).

\bibitem{miro} A.~E.~Miroshnichenko, S.~Flach andf Y.~S.~Kivshar, Rev. Mod. Phys. \textbf{82}, 2257 (2010).


\bibitem{sfan} S.~Fan, W.~Suh and J.~D.~Joannopoulos, 
%Temporal couple-mode theory for the Fano resonance in optical resonators,
J. Opt. Soc. Amm. A \textbf{20}, 569 (2003).

\bibitem{fan} S.~Fan and J.~D.~Joannopoulos, 
%Analysis of guided resonances in photonic crystal slabs,
Phys. Rev. B \textbf{65}, 235112 (2002).

\bibitem{astr} V.~N.~Astratov, I.~S.~Culshaw, R.~M.~Stevenson, D.~M.~Whittaker, M.~S.~Skolnick, T.~F.~Krauss  and R.~M.~De La Rue,
%Resonant Coupling of Near-Infrared Radiation to Photonic Band Structure Waveguides,
  J. Light. Technol. \textbf{17}, 2050 (1999).


\bibitem{mie}
G.~Mie, Ann. Physik \textbf{25}, 377 (1908).
\bibitem{hulst} H.~C.~van de Hulst, \textsl{Light scattering by small particles} Dover Publ. Inc, NY (1981)

\bibitem{trib} M.I.~Tribelsky, S.~Flach, A.~E.~Miroshnichenko, A.V.~Gorbach and Y.~S.~Kivshar, Phys. Rev. Lett. \textbf{100}, 04903 (2008).

\bibitem{rybin} M.~V.~Rybin, K.B.~Samusev, I.~S.~Sinev, G.~Semouchkin, E.~Semouchjina, Y.~S.~Kivshar, M.~F.~Limonov, Optics Express \textbf{21}, 30107 (2013);

\bibitem{rr} M.~V.~Rybin, D.~S.~Filonov, P.A.~Belov, Y.~S.~Kivshar and M.~F.~Limonov, Sci. Rep. \textbf{5}, 8774 (2015).

\bibitem{luk} B.~Lukyanchuk, N.~I. Zheludev, S.~A.~Maier, N.~J.~Halas, P.~Nordlander, H.~Giessen and Ch.~T.~Chong,
%The Fano resonance in plasmonic nanostructures and metamaterials,
  Nature Materials \textbf{9}, 707 (2010).

\bibitem{fano-nature} A.~N.~Poddubny, M.~V.~Rybin, M.~F.~Limonov and Y.~S.~Kivshar, 
%Fano interference governs wave transport in disordered systems,
Nature Communications \textbf{3}, 914 (2012).

\bibitem{io} K.~Inoue and K~Ohtaka,  in \cite{iok} p. 9. %\textsl{Survey of Fundamental Features of Photonic Crystals},


\bibitem{rrx} M.~V.~Rybin, A.~B.~Khanikaev, M.~Inoue, K.~B.~Samusev, M.~J.~Steel, G.~Yushin and M.~F.~Limonov, Phys. Rev. Lett. \textbf{103}, 023901 (2009). 



\bibitem{pendry} J.~B.~Pendry and A.~MacKinnon,
%Calculation of photonic dispersion relations,
 Phys. Rev. Lett. \textbf{69}, 2772 (1992).

\bibitem{pre} P. Marko\v{s}  and C.~M.~Soukoulis, Phys. Rev. E \textbf{65}, 036622 (2002).

\bibitem{stratton} J.~A.~Stratton,  \textsl{Electromagnetic Theory} (New York: Mc Graw-Hill Comp. 1941).

\bibitem{on} K.~Ohtaka, H.~Numata, Phys. Lett. \textbf{73A}, 411 (1979).

\bibitem{oua} K.~Ohtaka, T.Ueda and K.~Amemiya, Phys. Rev. B \textbf{57}, 2550 (1998).

\bibitem{pozn} To assure the numerical stability, Hankel functions are  normalized by the frequency-dependent 
function $H'_k(2\pi R/\lambda)$. 
Consequently, this function
scales coefficients $\beta_k$ shown in Figures throughout the paper.

\bibitem{pm-15} P.~Marko\v{s}, 
%Guided modes in photonic structures with left-handed components,
 ArXiv:1501.05125 (unpublished).


%\bibitem{kr} W. Kohn and N. Rostoker, Phys. Rev. \textbf{94}, 1111 (1954).

\bibitem{pc-asym} W.~M.~Robertson, G. Arjavalingam, R.~D.~Meade, K.~D.~Brommer, A.~M.~Rappe and J.~D.~Joannopoulos,
%Measurement of Photonic Band Structure in a Two-Dimensional Periodic Dielectric Array,
Phys. Rev. Lett. \textbf{68}, 2023  (1992).
\bibitem{sak} K.~Sakoda, Phys. Rev. B \textbf{52}, 7982 (1995).

\bibitem{sakoda-1997} K.~Sakoda,  
%Numerical analysis of the interference patterns in the optical transmission spectra of a square photonic lattice,
J. Opt. Soc. A. B \textbf{14}, 1961 (1997).

\bibitem{xx} Note that  the overlap frequency interval, found by the  
the transfer matrix  method in Fig. \ref{eps12-T}(c)
differs slightly from that calculated by the plane wave expansion (Fig. \ref{eps12-T}(a).
The origin of  this discrepancy lies in the numerical limitation  of the transfer matrix method
due to the  finite  discretization of space. 


\bibitem{as} M.~Abramowitz and I.~A.~Stegun,  \textsl{Handbook of Mathematical Functions} Dover Publ. (1965).

\bibitem{economou} C.~A.~Pfeiffer, E.~N.~Economou and K.~L.~Ngai, {Phys. Rev. B} \textbf{10}, 3038 (1974).

\bibitem{p2} Note that $\beta_2$ equals to scattering efficiency $Q_{\rm sca,2}$ \cite{rr} multiplied 
 by analytic contiguous function of the frequency.

\end{thebibliography}
\end{document}